\documentclass[conference]{IEEEtran}
\usepackage{todonotes}

\usepackage{amsmath,amssymb,bm}
\usepackage{rotating}
\usepackage{trfsigns}
\usepackage{aligned-overset}
\usepackage[sorting=none]{biblatex}

\AtEveryBibitem{
 \clearfield{url}
 \clearfield{issn}
}
\usepackage{siunitx}
\DeclareSIUnit{\ms}{\milli\second}   
\DeclareSIUnit{\us}{\micro\second}   
\DeclareSIUnit{\ns}{\nano\second}    
\DeclareSIUnit{\dBm}{dBm}  

\usepackage{comment}

\usepackage{subcaption}

\usepackage{trfsigns}
\usepackage[export]{adjustbox}

\usepackage{graphicx}
\usepackage{xcolor}

\usepackage{booktabs}    
\usepackage{colortbl}    
\usepackage[most]{tcolorbox} 

\usepackage{tikz}
\usetikzlibrary{shapes.geometric, arrows.meta, positioning,calc,backgrounds,shapes.multipart}
\usepackage{pgfplots}
\pgfplotsset{compat=1.18}

\usepackage{microtype}
\usepackage{placeins}

\usepackage[acronyms]{glossaries}
\newacronym{3gpp}{3GPP}{Third Generation Partnership Project}

\newacronym{admm}{ADMM}{Alternating Directions of Multipliers Method}
\newacronym{agv}{AGV}{Automated Guided Vehicels}
\newacronym{anm}{ANM}{Atomic Norm Minimization}
\newacronym{adc}{ADC}{Analog-to-Digital Converter}
\newacronym{acf}{ACF}{Auto-Correlation Function}
\newacronym{awgn}{AWGN}{Additive White Gaussian Noise}
\newacronym{asic}{ASIC}{Application Specific Integrated Circuit}
\newacronym{arpack}{ARPACK}{ARnoldi PACKage}
\newacronym{api}{API}{Application Programmable Interface}
\newacronym{aut}{AUT}{Antenna Under Test}
\newacronym[plural=AOI, firstplural=Areas of Interest (AOI)]{aoi}{AOI}{Area of Interest}
\newacronym{ai}{AI}{Artifical Intelligence}
\newacronym{aic}{AIC}{Akaike Information Criterion}
\newacronym{agc}{AGC}{Automatic Gain Control}
\newacronym{af}{AF}{Ambiguity Function}

\newacronym{blos}{BLOS}{Beyond Line Of Sight}
\newacronym{bp}{BP}{Basis Pursuit}
\newacronym{bs}{BS}{base station}
\newacronym{bpdn}{BPDN}{Basis Pursuit Denoising}
\newacronym{blue}{BLUE}{Best Linear Unbiased Estimator}
\newacronym{bic}{BIC}{Bayesian Information Criterion}
\newacronym{bibo}{BIBO}{Bounded-Input-Bounded-Output}
\newacronym{btp}{BTP}{Bistatic Target Path}
\newacronym{bt}{BT}{bandwidth-time}

\newacronym{ci}{CI}{critical infrastructure}
\newacronym{cnn}{CNN}{Convolutional Neural Network}
\newacronym{crlb}{CRLB}{Cramér-Rao lower bound}
\newacronym{cs}{CS}{Compressed Sensing}
\newacronym{cf}{CF}{Crest factor}
\newacronym{cr}{CR}{Compression Ratio}
\newacronym{cu}{CU}{Central Unit}
\newacronym{cmos}{CMOS}{Complementary Metal Oxide Semiconductor}
\newacronym{cots}{COTS}{Commercial-Off-The-Shelf}
\newacronym{cpu}{CPU}{Central Processing Unit}
\newacronym{cfar}{CFAR}{Constant False Alarm Rate}
\newacronym{cvx}{CVX}{Convex Optimization toolboX}
\newacronym{comp}{CoMP}{Cooperative Multi-Point}
\newacronym{cpcl}{CPCL}{Cooperative Passive Coherent Location}
\newacronym{csi}{CSI}{channel state information}
\newacronym{cir}{CIR}{Channel Impulse Response}
\newacronym{cdf}{CDF}{Cumulative Distribution Function}
\newacronym{cam}{CAM}{Cooperative Awareness Messages}
\newacronym{cpx}{CPX}{Cyclic Prefix}

\newacronym{doa}{DoA}{direction of arrival}
\newacronym{dod}{DoD}{Direction of Departure}
\newacronym[plural=DMC,firstplural=Diffuse Multipath Components (DMC)]{dmc}{DMC}{Diffuse Multipath Components}
\newacronym{das}{DAS}{Delay-and-Sum}
\newacronym{dft}{DFT}{Discrete Fourier Transform}
\newacronym{idft}{iDFT}{Inverse Discrete Fourier Transform}
\newacronym{dtft}{DTFT}{Discrete Time Fourier Transform}
\newacronym{idtft}{iDTFT}{Inverse Discrete Time Fourier Transform}
\newacronym{dct}{DCT}{Discrete Cosine Transform}
\newacronym{dsp}{DSP}{Digital Signal Processor}
\newacronym{dnn}{DNN}{Deep Neural Network}
\newacronym{dlr}{DL}{Deep Learning}
\newacronym{dbs}{DBS}{Distributed Base Station}
\newacronym{dmrs}{dmrs}{DeModulation Reference Signal}

\newacronym{eadf}{EADF}{Effective Aperture Distribution Function}
\newacronym{eirp}{EIRP}{equivalent isotropically radiated power}
\newacronym{etadf}{ETADF}{Effective Time-Aperture Distribution Function}
\newacronym{esprit}{ESPRIT}{Estimation of Signal Parameters via Rotational Invariance Techniques}
\newacronym{ett}{ETT}{Eigenvalue Threshold Test}
\newacronym{edc}{EDC}{Efficient Detection Criterion}
\newacronym{eft}{EFT}{Exponential Fitting Test}
\newacronym{expm}{EM}{Expectation Maximization}
\newacronym{ekf}{EKF}{Extended Kalman Filter}
\newacronym{ems}{FG EMS}{Fachgebiet Elektronische Messtechnik und Signalverarbeitung}
\newacronym{etof}{eToF}{excess time of flight}
\newacronym{etoa}{eToA}{excess time of arrival}
\newacronym{edoppler}{eDoppler}{excess Doppler}
\newacronym{evv}{eVV}{excess velocity vector}

\newacronym{fc}{FC}{fully-connected}
\newacronym{ft}{FT}{Fourier Transform}
\newacronym{fht}{FHT}{Fast Hadamard Transform}
\newacronym[longplural={Fast Fourier Transforms}]{fft}{FFT}{Fast Fourier Transform}
\newacronym{fmcw}{FMCW}{Frequency-Modulated Continuous-Wave}
\newacronym{fpga}{FPGA}{Field Programmable Gate Array}
\newacronym{fri}{FRI}{Finite Rate of Innovation}
\newacronym{fir}{FIR}{Finite Impulse Response}
\newacronym{fim}{FIM}{Fisher Information Matrix}
\newacronym{fmc}{FMC}{Full Matrix Capture}
\newacronym{fista}{FISTA}{Fast Iterative Shrinkage-Thresholding Algorithm}
\newacronym{frvm}{FRVM}{Fast Relevance Vector Machine}
\newacronym{flop}{FLOP}{Floating Point Operation}
\newacronym{fl}{FL}{Federated Learning}
\newacronym{frf}{FRF}{Frequency Response Functions}
\newacronym{fr1}{FR1}{frequency range 1}
\newacronym{fr2}{FR2}{frequency range 2}
\newacronym{fr3}{FR3}{frequency range 3}
\newacronym{fdma}{FDMA}{Frequency-Division Multiple Access}

\newacronym{gfcs}{grid-free CS}{grid-free compressive sensing}
\newacronym{gpu}{GPU}{Graphical Processing Unit}
\newacronym{gtd}{GTD}{Geometrical Theory of Diffraction}
\newacronym{gan}{GAN}{Generative Adversarial Network}
\newacronym{gdop}{GDoP}{geometric dilution of precision}
\newacronym{gnb}{gNB}{next generation Node B}
\newacronym{gpsdo}{GPSDO}{GPS disciplined oscillators}

\newacronym{hdop}{HDoP}{horizontal dilution of precision}
\newacronym{hrpe}{HRPE}{High Resolution Parameter Estimator}
\newacronym{hfss}{Ansys HFSS}{Ansys High Frequency Electromagnetic Simulation Software}

\newacronym[longplural={Inverse Fast Fourier Transforms}]{ifft}{IFFT}{Inverse Fast Fourier Transform}
\newacronym{ir}{IR}{Impulse Response}
\newacronym{iid}{iid}{independent and identically distributed}
\newacronym{iir}{IIR}{Infinite Impulse Response}
\newacronym{irf}{IRF}{Impulse Response Function}
\newacronym{icas}{ICAS}{Integrated Communications and Sensing}
\newacronym{isac}{ISAC}{Integrated Sensing and Communications}
\newacronym{ista}{ISTA}{Iterative Shrinkage-Thresholding Algorithm}
\newacronym{ici}{ICI}{Inter Carrier Interference}
\newacronym{iot}{IoT}{Internet of Things}

\newacronym{jt}{JT}{Joint Transmission}
\newacronym{jpda}{JPDA}{Joint Probabilistic Data Association}

\newacronym{kld}{KLD}{Kullback-Leibler Divergence}
\newacronym{kpi}{KPI}{Key Performance Indicator}

\newacronym{ls}{LS}{Least Squares}
\newacronym{lasso}{LASSO}{Least Absolute Shrinkage and Selection Operator}
\newacronym{lse}{LSE}{Line Spectral Estimation}
\newacronym{lfsr}{LFSR}{Linear Feedback Shift Register}
\newacronym{lms}{LMS}{least-mean-squares}
\newacronym{lo}{LO}{Local Oscillator}
\newacronym{los}{LOS}{line of sight}
\newacronym{lti}{LTI}{Linear Time-Invariant}
\newacronym{ltv}{LTV}{Linear Time-Variant}
\newacronym{lam}{LAM}{Large Area Monitoring}

\newacronym{mimo}{MIMO}{Multiple Input Multiple Output}
\newacronym{mmv}{MMV}{multiple measurement vectors}
\newacronym{mmse}{MMSE}{misspecified mean squared error}
\newacronym{mse}{MSE}{mean squared error}
\newacronym{bce}{BCE}{Binary Crossentropy}
\newacronym{mlbs}{MLBS}{Maximum Length Binary Sequence}
\newacronym{mno}{MNO}{Mobile Network Operator}
\newacronym{mpc}{MPC}{Multipath Component}
\newacronym{msm}{MSM}{M-Sequence Method}
\newacronym{mwc}{MWC}{Modulated Wideband Converter}
\newacronym{mpm}{MPM}{Matrix Pencil Method}
\newacronym{mpu}{MPU}{Microprocessor Unit}
\newacronym{mumimo}{MU MIMO}{multi-user MIMO}
\newacronym{mu}{MU}{multi-user}
\newacronym{ms}{MS}{multi-sensor}
\newacronym{msue}{MS\textsubscript{UE}}{multi-sensor using user equipment}
\newacronym{msru}{MS\textsubscript{RU}}{multi-sensor using radio unit}
\newacronym{ml}{ML}{Machine Learning}
\newacronym{mri}{MRI}{Magnetic resonance imaging}
\newacronym{music}{MUSIC}{Multiple Signal Classification}
\newacronym{mkl}{MKL}{Math Kernel Library}
\newacronym{mcrb}{MCRB}{Misspecified Cramér-Rao Bound}
\newacronym{mmle}{MMLE}{Misspecified Maximum-Likelihood Estimator}
\newacronym{mbpe}{MBPE}{Model-Based Propagation Parameter Estimation}
\newacronym{mrf}{MRF}{Multistatic Reflectivity Function}
\newacronym{mec}{MEC}{Mobile Edge Cloud}
\newacronym{msisac}{MS-ISAC}{Multi-Sensor ISAC}
\newacronym{mht}{MHT}{Multi-Hypotheses Tracker}
\newacronym{mi}{MI}{Mutual Information}

\newacronym{ndt}{NDT}{Nondestructive Testing}
\newacronym{nde}{NDE}{Nondestructive Evaluation}
\newacronym{nn}{NN}{Neural Net}
\newacronym{nist}{NIST}{National Institute of Standards and Technology}
\newacronym{npn}{NPN}{Non-Public Networks}
\newacronym{nr}{NR}{New Radio}
\newacronym{nlos}{NLOS}{Non Line of Sight}
 
\newacronym{omp}{OMP}{Orthogonal Matching Pursuit}
\newacronym{oop}{OOP}{Object Oriented Programming}
\newacronym{ota}{OTA}{Over The Air}
\newacronym{ofdm}{OFDM}{orthogonal frequency-division multiplexing}
\newacronym{ofdma}{OFDMA}{Orthogonal Frequency-Division Multiple Access}
\newacronym{afdm}{AFDM}{Affine Frequency Division Multiplexing}
\newacronym{otfs}{OTFS}{Orthogonal Time Frequency Space}

\newacronym{pdp}{PDP}{Power Delay Profile}
\newacronym{pap}{PAP}{Power Angular Profile}
\newacronym{pn}{PN}{Pseudo-Noise}
\newacronym{pwc}{PWC}{Plane Wave Compounding}
\newacronym{pcl}{PCL}{Passive Coherent Location}
\newacronym{pdsch}{PDSCH}{physical downlink shared channel}
\newacronym{pwi}{PWI}{Plane Wave Imaging}
\newacronym{pura}{PURA}{Patch Uniform Rectangular Array}
\newacronym{pymax}{PyMAX}{Python Maximization Approach}
\newacronym{pts}{PTS}{Pseudo-True Solution}
\newacronym{pdf}{pdf}{probability density function}
\newacronym{pi}{PI}{Principal Investigator}
\newacronym{pidl}{PIDL}{Physics Informed Deep Learning}
\newacronym{pinn}{PINN}{Physics Informed Neural Network}
\newacronym{pbdl}{PBDL}{Physics Based Deep Learning}
\newacronym{prs}{PRS}{position reference signal}
\newacronym{ps}{PS}{protected site}

\newacronym{qam}{QAM}{quadrature amplitude modulation}

\newacronym{ran}{RAN}{radio access network}
\newacronym{ranic}{RIC}{RAN Intelligent Controller}
\newacronym{relu}{ReLU}{Rectified Linear Unit}
\newacronym{resnet}{ResNet}{Residual Neural Network}
\newacronym{ram}{RAM}{Random Access Memory}
\newacronym{rcs}{RCS}{Radar Cross Section}
\newacronym{rd}{RD}{Random Demodulator}
\newacronym{rx}{Rx}{receiver}
\newacronym{rem}{REM}{Reconstruction Error Metric}
\newacronym{rmse}{RMSE}{Root Mean Squared Error}
\newacronym{rms}{RMS}{root mean squared}
\newacronym{ric}{RIC}{Restricted Isometry Constant}
\newacronym{rip}{RIP}{Restricted Isometry Property}
\newacronym{ris}{RIS}{Reconfigurable Intelligent Surface}
\newacronym{rc}{RC}{Raised Cosine}
\newacronym{roi}{ROI}{Region of Interest}
\newacronym{roc}{ROC}{Region of Convergence}
\newacronym{rt}{RT}{Raytracing}
\newacronym{rimax}{RIMAX}{Richter Maximization Approach}
\glsunset{rimax}
\newacronym{rvm}{RVM}{Relevance Vector Machine}
\newacronym{rss}{RSS}{Received Signal Strength}
\newacronym{rfid}{RFID}{Radio Frequency Identification}
\newacronym{refodat}{REFODAT}{Repositorium für Forschungsdaten in Thüringen}
\newacronym{re}{RE}{Resource Element}
\newacronym{rb}{RB}{Resource Block}
\newacronym{rrh}{RRH}{Remote Radio Head}
\newacronym{rru}{RRU}{Remote Radio Unit}
\newacronym{ru}{RU}{radio unit}
\newacronym{du}{DU}{distributed unit}
\newacronym{bbu}{BBU}{baseband unit}
\newacronym{openran}{O-RAN}{open radio access network}

\newacronym{samurai}{SAMURAI}{Synthetic Aperture Measurements of Uncertainty in Angle of Incidence}
\newacronym[plural=SC,firstplural=Specular Components (SC)]{sc}{SC}{Specular Components}
\newacronym{sdp}{SDP}{semi-definite program}
\newacronym{sdr}{SDR}{Signal to Diffuse Ratio}
\newacronym{simd}{SIMD}{Single Instruction Multiple Data}
\newacronym{svd}{SVD}{singular value decomposition}
\newacronym{svm}{SVM}{Support Vector Machine}
\newacronym{soe}{SOE}{Sparsity Order Estimation}
\newacronym{sgd}{SGD}{Stochastic Gradient Descent}
\newacronym{stuca}{StUCA}{Stacked Uniform Circular Array}
\newacronym{spucpa}{SPUCPA}{Stacked Polarimetric Uniform Circular Patch Array}
\newacronym{suca}{SUCA}{Stacked Uniform Circular Array}
\newacronym{saft}{SAFT}{Synthetic Aperture Focusing Technique}
\newacronym{sota}{SOTA}{State of the Art}
\newacronym{ssd}{SSD}{Solid State Device}
\newacronym{ssr}{SSR}{Sparse Signal Recovery}
\newacronym{sa}{SA}{Synthetic Aperture}
\newacronym{sh}{SH}{Spherical Harmonics}
\newacronym{spw}{SPW}{Single Plane Wave}
\newacronym{shm}{SHM}{Structural Health Monitoring}
\newacronym{snr}{SNR}{signal-to-noise ratio}
\newacronym{stela}{STELA}{Soft-Thresholding with Exact Line Search Algorithm}
\newacronym{siso}{SISO}{Single Input Single Output}
\newacronym{simo}{SIMO}{Single Input Multiple Output}
\newacronym{swe}{SWE}{Spherical Wave Expansion}
\newacronym{sme}{SME}{Spherical Mode Expansion}
\newacronym{sage}{SAGE}{Space-Alternating Generalized Expectation-Maximization}
\newacronym{stft}{STFT}{Short Time Fourier Transformation}
\newacronym{sl}{SL}{Side Link}
\newacronym{slc}{SLC}{Sensor Level Cooperation}
\newacronym{scf}{ScF}{Scattering Function}
\newacronym{sf}{SF}{Spreading Function}
\newacronym{sar}{SAR}{Synthetic Aperture Radar}
\newacronym{ss}{SS}{single station}

\newacronym{th}{T\&H}{Track and Hold}
\newacronym{tf}{TF}{Transfer Function}
\newacronym{tx}{Tx}{transmitter}
\newacronym{twista}{TWISTA}{Two-step Iterative Shrinkage-Thresholding Algorithm}
\newacronym{tof}{ToF}{time of flight}
\newacronym{tdoa}{TDoA}{Time Difference of Arrival}
\newacronym{toa}{ToA}{Time of Arrival}
\newacronym{tdd}{TDD}{Time Division Duplex}
\newacronym{tdma}{TDMA}{Time-division Multiple Access}
\newacronym{trf}{TRF}{Time Reversal Focusing}
\newacronym{tr}{TR}{Time Reversal}

\newacronym{uca}{UCA}{uniform circular array}
\newacronym{ura}{URA}{uniform rectangular array}
\newacronym{ula}{ULA}{Uniform Linear Array}
\newacronym{uwb}{UWB}{Ultra-Wideband}
\newacronym{usndt}{US-NDT}{Ultrasonic Non-destructive Testing}
\newacronym{ue}{UE}{user equipment}
\newacronym{ul}{UL}{Uplink}
\newacronym{dl}{DL}{Downlink}
\newacronym{uav}{UAV}{unmanned aerial vehicle}
\newacronym{udc}{UDC}{Up/Down Converter}
\newacronym{utm}{UTM}{uncrewed aircraft systems traffic management}
\newacronym{usrp}{USRP}{Universal Software Radio Peripheral}

\newacronym{vna}{VNA}{Vector Network Analyser}
\newacronym{vsh}{VSH}{Vector Spherical Harmonics}
\usepackage[
    draft,
]{hyperref}

\usepackage{orcidlink}
\usepackage{cleveref}

\crefname{figure}{Fig.}{Fig.}
\Crefname{figure}{Fig.}{Fig.}

\title{Multi-Sensor Integrated Sensing and Communication for Critical Infrastructure Protection}

\author{
    \IEEEauthorblockN{
        Reiner Thom\"a\IEEEauthorrefmark{1}\,\orcidlink{0000-0002-9254-814X},
  	    Gerd Sommerkorn\IEEEauthorrefmark{1}\,\orcidlink{0009-0003-1111-322X},
        Christian Schneider\IEEEauthorrefmark{1}\IEEEauthorrefmark{2}\,\orcidlink{0000-0003-1833-4562},
        Thomas Dallmann\IEEEauthorrefmark{1}\IEEEauthorrefmark{2}\,\orcidlink{0000-0003-4655-6568},
	}                                     
\\
	\IEEEauthorblockA{
		\IEEEauthorrefmark{1}Technische Universit\"at Ilmenau, Institute for Information Technology, Ilmenau, Germany\\
		\IEEEauthorrefmark{2}Fraunhofer Institute of Integrated Circuits, Dep. EMS, Ilmenau, Germany\\
	}
}

\begin{document}
\maketitle
\thispagestyle{empty} 
\pagestyle{empty}     

\begin{abstract}
\Gls{isac} will become a service in future mobile communication networks.
It enables the detection and recognition of passive objects and environments using radar-like sensing.
One promising first application is the protection of \gls{ci}, for example by monitoring the lower airspace above sensitive sites or facilities to prevent unauthorized drone overflights.
Our proposal is based on the concept of a distributed \gls{ms}-\gls{isac}.
We assume deploying three or more additional passive sniffing sensors near the \gls{ps} of a \gls{ci}.
The sniffers are connected via \gls{dl} / \gls{ul} to the distant illumination \gls{bs}.
Multistatic range-Doppler estimation, including synchronization, is performed according to the \gls{cpcl} principle.
The multistatic architecture has several advantages over the often considered quasi-monostatic architecture where one sniffer is located close to the base station.
We discuss the advantages and disadvantages of both approaches and compare their performance for the considered use case in terms of coverage and \gls{gdop}.
\end{abstract}


\section{Introduction and Motivation}
\label{sec:introduction}
The use of \gls{uav}, also known as drones, is becoming increasingly popular not only among hobbyists but also for commercial applications in civil engineering, agriculture and forestry, logistics, search and rescue, etc.
\Glspl{mno} are preparing to support \gls{utm} within the U-Space framework, the lower airspace designated and regulated for \glspl{uav}. Similar to manned civil aviation, ensuring the safety of air traffic in U-Space requires an independent monitoring system that detects flying objects that, for various reasons, do not comply with applicable regulations.
Regardless of this, a potential threat has recently emerged that involves deliberate, unauthorized drone overflights.
It can be assumed that these drone activities may be intended for provocation, espionage or even sabotage.
Properties that are part of \gls{ci} are particularly vulnerable to this.
In addition to large industrial grounds, logistics centers, and national defense sites, energy, communications, and transportation infrastructure facilities are particularly at risk, as they are often relatively unprotected and distributed across a wide rural area.
This results in a latent threat scenario that conflicts with the need for regulated and safe commercial use of U-Space.
While remote-controlled drones can be easily detected via radio surveillance due to their radio emissions, malicious drones conceal their presence by not transmitting any active signals.
Passive drones can only be detected by radar systems when the target area is illuminated by a radio transmitter.

However, area-wide radar monitoring of U-space is not viable by existing governmental air-traffic control because of several reasons, including limited radar coverage, regulatory issues, technical, and economic effort.
What we need is a ubiquitous, scalable, and cost-effective sensor system for monitoring the lower airspace that aggregates a nationwide situational picture.
It should enable the integration of additional sensors and external information, generate a situation report as needed, and make this report available to the relevant security authorities and operators of critical infrastructure for decision-making purposes.
The challenges are manifold.
For instance, there are not enough radio frequency bands available for dedicated radar sensors.
In contrast, the seamless integration of radar sensing into existing mobile networks promises a sustainable solution that conserves resources by reusing existing infrastructure and allocated frequency bands.
This is known as \gls{isac}.
In addition to a comprehensive radio interface for sensing, \gls{isac} provides the entire network infrastructure which is used for sensor control, data transport and fusion including edge processing.
The actual local \gls{isac} data can be linked with data from other sensors and fused between sensor clusters.
The \gls{mno} takes responsibility for administering the \gls{isac} sensing quality on the cluster or network level (``Sensing as a Service'').
This service is made available to various users.
Eventually a comprehensive situational picture is aggregated.
To this end, the compressed information from various local \gls{isac} clusters of different operators is aggregated, and, where applicable, combined with information from other sources.
The service at this level is tailored to the requirements of the respective client.

Even though \gls{isac} may have many facets and standardization is still underway (3GPP, ETSI), in this article, we will propose a pragmatic approach that enables \glspl{isac} for the use case of \gls{ci} protection briefly described above; see also \cite{Shatov2024JRC}.
Our proposal relies exclusively on existing 4G/5G radio signaling schemes and avoids making any changes to the hardware of the \glspl{bs} or to their internal \gls{ran} protocols.
We recommend installing three or more additional sniffing sensors near the \gls{ps}.
These sniffers act as distributed \gls{ms}-\gls{isac} \cite{Thoma2026MSISAC} sensors and are connected to the \gls{bs} as standard \gls{ue} devices.
This is basically different from other proposals (e.g. \cite{Saur2026UAVDetection}) where a sniffer is deployed close to the \gls{bs} to mimic a monostatic radar.
We will demonstrate that our setup has fundamental advantages in terms of coverage and overall performance.

\section{Proposed MS ISAC Architecture Setup}
\label{sec:architectures}
\begin{figure}[b]
    \centering
    \includegraphics[width=.9\linewidth,bgcolor=gray!10,rndcorners=5,rndframe={color=gray!50, width=\fboxrule, sep=\fboxsep}{5}]{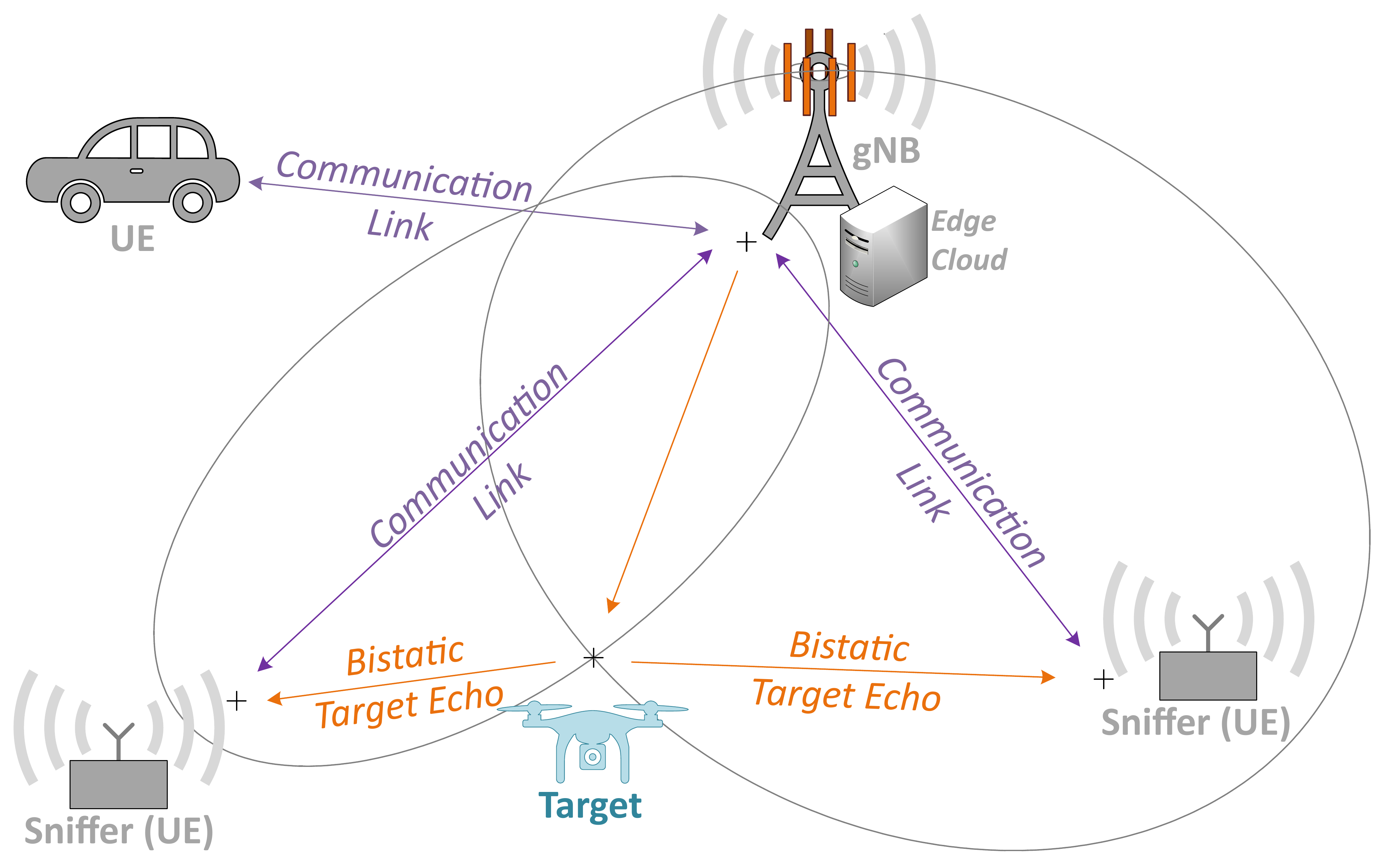}
    \caption{Generic MS-ISAC architecture. The two sniffers are connected to the \gls{gnb} by the regular \gls{ul} / \gls{dl}. The estimated \gls{etof} are indicated as ellipses. The crossing point is the target position.}
    \label{fig:architecture}
    \vspace{-15pt}
\end{figure}

Obviously, the current discussion at 3GPP about \gls{isac} architectures still seems to focus on the monostatic case for infrastructure centric \gls{isac} \cite{Blandino2026MonoStaSenUAV}.
As true full duplex radio access is not in reach, quasi-monostatic access is considered, where an auxiliary sniffer device close to the \gls{bs} is used  \cite{Saur2026UAVDetection}.
Apart from the strong direct feed trough (self-interference), there are several drawbacks of this setup.
The first problem concerns target localization estimation, which, in the case of a \gls{ss}-\gls{isac}, with co-located transmit and receive antennas, requires not only range estimate (distance to \gls{gnb}) but also an estimate of the direction of arrival.
However, common \gls{gnb} array access and signal processing is not very well suited for \gls{doa} estimation.
It is more suitable for beamforming (spatial precoding), which allows only low resolution \gls{doa} estimation, e.g. \qtyrange[range-phrase={ to ca. }]{12.7}{25}{\degree} beamwidth at the sector corners (because of beam squint) for an 8 element \gls{ura} in azimuth.
Consequently, the cross-range resolution and accuracy decreases as the distance from the target increases.
That does not rule out the possibility of greater \gls{doa} estimation accuracy in a high \gls{snr} regime with proper off-grid beampattern interpolation.
But because of the radar equation, \gls{snr} decreases with $R^{-4}$, where $R$ is the radial distance to the target.
Therefore, monostatic \gls{ss}-\gls{isac} has often an unsatisfactory geolocation performance.
\cite{Saur2026UAVDetection} does not even discuss \gls{doa} estimation, but only range-Doppler estimation, which may be sufficient for target detection in \gls{ss}-\gls{isac}, but not for target localization. 

True multi-target high resolution \gls{doa} estimation is possible only with mutual antenna correlation matrix subspace processing (\gls{esprit}, \gls{music}) or with model-based maximum likelihood methods.
Both are not directly compatible with standard spatial precoding processing (see \cite{Thoma2026MSISAC} for high resolution \gls{doa} estimation with spatial precoding).
Therefore, often only beam-specific range-Doppler estimation is considered \cite{Keskin2025FundTradoffMonoStaISAC}.
Another limitation is that a monostatic single station radar can only estimate the radial speed of the target since there is no cross-range Doppler shift measurement available.
This is because of a lack of geometric degrees of freedom.
Moreover, a monostatic radar cannot deliver any diversity gain for more reliable detection (see the ''stealth” problem).
Multiple monostatic sensing could help a bit.
However, the cellular network is not designed for overlapping radio access.
Therefore, it is highly unlikely that a target would be visible at the same time from multiple macro \glspl{bs} belonging to the same \gls{mno}, perhaps except from target locations at the cell borders.
However, these cases are rare as resources of neighboring \glspl{gnb} are allocated towards mutual interference mitigation. 

Distributed \gls{ms}-\gls{isac} potentially avoids all the disadvantages mentioned above for \gls{ss}-\gls{isac}.
The design principles and advantages of \gls{ms}-\gls{isac} were comprehensively worked out in \cite{Thoma2026MSISAC}.
According to the prospective use case of critical infrastructure protection, we will focus on infrastructure centric \gls{isac}.
This means that only the \gls{ran} infrastructure and some dedicated auxiliary \gls{ue} equipment that is registered and qualified for the this specific sensing service is involved in the sensing cycle.
Third party communication \glspl{ue} are not directly included, except that the communication \gls{dl} to some contractual user is reused (required) for scene illumination. 

There are several approaches for distributed \gls{ms}-\gls{isac}.
Here we assume multiple sniffers acting as remote (hence bistatic) sensors.
These sniffers are deployed as dedicated devices at strategically selected locations on behalf of the mobile network operator.
They are connected as normal \glspl{ue} to the \gls{gnb} via \gls{ul} and \gls{dl}, see \cref{fig:architecture}.
We refer to this approach as \gls{msue}-\gls{isac}.
The \gls{ul} / \gls{dl} connection already ensures \gls{rx}-\gls{tx} frame synchronization.
The \gls{ue} sniffers receive the \gls{dl} and perform sensing in the \gls{msue}-\gls{isac} broadcast mode using the full multiuser \gls{pdsch} data stream together with all reference signals transmitted in \gls{dl}.
This means that all 5G \gls{nr} signaling, scheduling, and precoding functions intended for communication can also be used for \gls{isac}, e.g. to use reference and pilot signals for sensing channel state estimation, equalization and link adaptation.
The idea behind is the \gls{cpcl} principle, also known as cooperative passive radar described at first in \cite{Thoma2019CPCL}.
With \gls{cpcl} we can make sure that the full \gls{pdsch} communication data stream is used for bistatic delay-Doppler estimation at the \gls{rx}.
This explains why the frequency resources allocated for communication are reused entirely for radar sensing, thereby maximizing the \gls{snr} compared to \gls{isac} approaches that use only reference and pilot signals.
Also, \glspl{prs} can be activated in \gls{dl} if there is not enough communication capacity requested.
This even works in the multiuser downlink, since like in passive radar the sensor does not distinguish with respect to the dedicated receiver.
Only the transmitted waveform, not the authorized data content, must be recovered at the \gls{ue} to be used as correlation reference to calculate \gls{etof} and \gls{edoppler}.
First single link target related location parameters are calculated at the \gls{ue} and transmitted via the \gls{ul} for fusion of data collected from all sniffers at the \gls{gnb} on the application level.
Local estimation at the \gls{ue} is import for data reduction.
The procedures must consider that joint target detection, multitarget data association and 3D estimation of the dynamic target state vector in the fusion center at the \gls{gnb} becomes possible.  

Key advantages of \gls{ms}-\gls{isac} over the (quasi) monostatic \gls{ss}-\gls{isac} are:
(i) A full 3D dynamic target state estimation is possible through multilateration (bistatic distance and Doppler) and thus without \gls{doa} estimation;
(ii) A suitable distribution of sniffer positions around the target area not only enables 3D multilateration but also significantly improves the radar link budget.
This is particularly true when the area to be observed is clearly defined and not too large, and when the illuminating macro \gls{gnb} is not too close.
This is a situation which is very typical for critical infrastructure protection in rural areas. An excess number of sniffers allows further reduction of estimation variance and can detect outliers the may happen if the line of sight to the target is coincidentally obstructed;
(iii) In the specific geographical situation described, it is probable that the dedicated \gls{ue} sniffer devices are served by the same DL transmission beam, and this beam also illuminates the targets.
Therefore, we don’t need target specific spatial precoding since the regular precoding requested by the sniffer \gls{ue} is sufficient. 

Another key advantage of the \gls{cpcl} principle is that \gls{etoa} and \gls{edoppler} measurements reduce to differential coherent methods since excess target range and Doppler are measured relative to direct \gls{tx} to \gls{rx} \gls{los} from the same impulse response.
This largely relaxes transmit-receive synchronization \cite{Thoma2026MSISAC}.
\Gls{etoa} and \gls{edoppler} measurements are based on correlation, which requires knowledge of the transmitted waveform used as correlation reference at the receiver.
Although standard \gls{ofdm} equalization already provides an initial estimate of the transmitted waveform, the result is not sufficient for \gls{isac}.
Reasons are twofold: The dynamic range requirements for \gls{isac} are generally higher than those for communication.
The error is due to noise and poor interpolation from sparse \gls{csi} estimates.
Another reason may be nonlinear distortion at the power amplifier which is not equalized.
To enhance the quality of correlation reference recovery, we proposed a concatenated two step retrieval procedure (turbo \gls{cpcl}, \cite{Thoma2026MSISAC}) that starts with standard \gls{csi} based equalization for first transmit signal recovery and continues with a second step that uses the complete \gls{pdsch} communication data estimated in the first step.
This extended equalization is also necessary because the \gls{ms}-\gls{isac} broadcast mode requires equalization of the entire multiuser downlink which is not equalized in a standard communication receiver.

\section{Realistic Deployment Example}
\label{sec:deployment}
\begin{figure}[t]
    \centering
    \includegraphics[width=.9\linewidth,trim=0 0 2 2, clip,bgcolor=gray!10,rndcorners=5,rndframe={color=gray!50, width=\fboxrule, sep=\fboxsep}{5}]{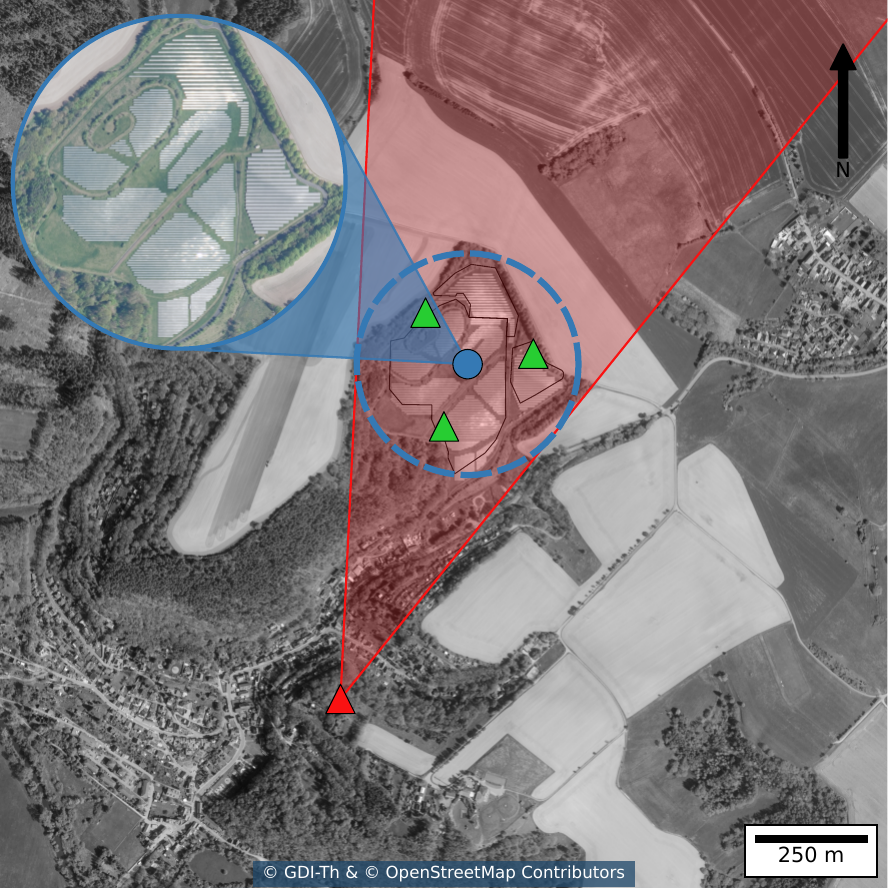}
    \caption{Example deployment scenario consisting of a solar field that represents the \gls{ps} and a remote macro \gls{bs} at an elevated position on a hill near Bad Salzungen, Germany. The three hypothetic sniffer \glspl{ue} are indicated as green triangles, the \gls{bs} is the red triangle.}
    \label{fig:deployment}
    \vspace{-15pt}
\end{figure}

We have chosen a realistic example scenario as given in \cref{fig:deployment}.
The \gls{ps} is a solar field, indicated by the blue dashed circle, with an approximate diameter of $\SI{500}{\metre}$.
We assume three sniffer \glspl{ue} located at the perimeter of the solar field.
The elevated BS is $\approx\SI{800}{\metre}$ away (relative to the median of the sniffers positions).

The \gls{bs} antenna and the solar array are located at roughly the same height.
With the regular elevation beamwidth, it should be no problem to illuminate the lower airspace over the specified distance.
\Cref{fig:deployment} also shows the \SI{3}{\dB} azimuth contour of the \gls{dl} transmission beam that is requested by the three \gls{ue} sniffers.
It appears that the target is also illuminated by the same beam.
Since the \gls{bs}’s coverage area extends beyond the target area, there may be further communication users located behind the solar array hosted by the same beam who request \gls{dl} data traffic which helps to better illuminate the target.

\section{MS ISAC Coverage Analysis}
\label{sec:analysis}
\begin{figure*}[t]
    \centering
    \begin{minipage}{0.48\textwidth}
        \centering
        \includegraphics[width=.9\linewidth,trim=0 0 2 2, clip,bgcolor=gray!10,rndcorners=5,rndframe={color=gray!50, width=\fboxrule, sep=\fboxsep}{5}]{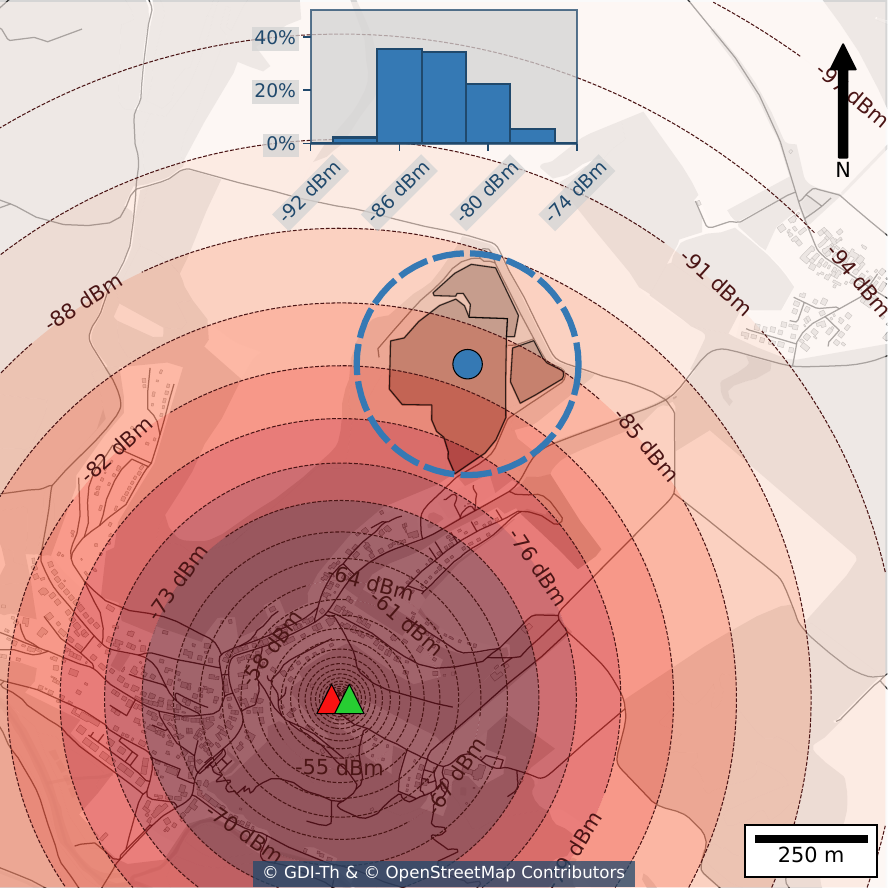}
        \caption*{a) Monostatic case}
    \end{minipage}
    \hfill
    \begin{minipage}{0.48\textwidth}
        \centering
        \includegraphics[width=.9\linewidth,trim=0 0 2 2, clip,bgcolor=gray!10,rndcorners=5,rndframe={color=gray!50, width=\fboxrule, sep=\fboxsep}{5}]{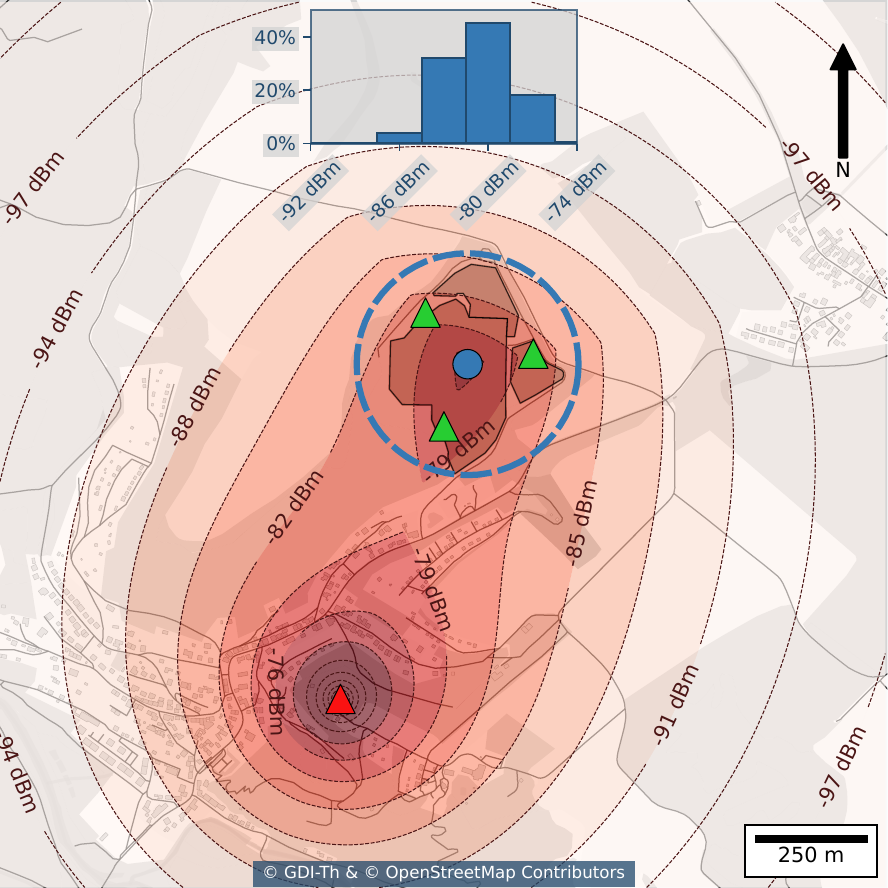}
        \caption*{b) Multistatic case}
        \end{minipage}
    \vspace{-3pt}
    \caption{Spatial distribution of the simulated received power \(P_{\mathrm{r}}\) within the deployment scenario (represented by isolines), with the corresponding power distribution across the \gls{ps} illustrated as an inset histogram.}
    \label{fig:recpower}
    \vspace{-15pt}
\end{figure*}

Radar system performance depends on the received power \(P_{\mathrm{r}}\) reflected from the target \(\mathrm{s}\) as given in the logarithmic form for the general bistatic configuration:

\begin{equation}
\begin{aligned}
P_{\mathrm{r}} =\;& P_{\mathrm{t}} + G_{\mathrm{t}} + G_{\mathrm{r}} + \sigma_{\mathrm{s}} + 20\log_{10}(\lambda)- 30\log_{10}(4\pi)\\
&- 20\log_{10}(R_{\mathrm{t\rightarrow s}}) - 20\log_{10}(R_{\mathrm{s\rightarrow r}}).
\end{aligned}
\label{eq:bistatic_radar}
\end{equation}

where \(P_{\mathrm{t}}\) is the transmitted power (in \unit{\dBm}), \(G_{\mathrm{t}}\) and \(G_{\mathrm{r}}\) are the transmitter and receiver antenna gains (in \unit{\dB i}), $\lambda$ is the wavelength (in \unit{\metre}),  \(R_{\mathrm{t\rightarrow s}}\) and \(R_{\mathrm{s\rightarrow r}}\) and  are the transmitter-to-target and target-to-receiver distances (in \unit{\metre}), respectively. \(\sigma_{\mathrm{s}}\) denotes the bistatic \gls{rcs} that describes the fraction of power scattered from the target to the receiver within one resolved range bin. Here it is given in \unit{\dB s\metre}. The \gls{rcs} must be regarded as a random variable whose expected value depends on the size, material, and orientation of the target and whose variance is related to the structure of the target. Moreover, it depends on the polarization and angles of the impinging and outgoing waves.
 \Cref{eq:bistatic_radar} considers only propagation in free space routed over the target, without the \gls{los} between the transmitter and the receiver, and without clutter. By introducing the range ratio
\begin{equation}
    a = \frac{R_{\mathrm{t\rightarrow s}}}{R_{\mathrm{s\rightarrow r}}},
    \label{eq:rangeratio}
\end{equation}
the independent distance terms in \cref{eq:bistatic_radar} can be unified. 
\begin{equation}
\begin{aligned}
P_{\mathrm{r}} =\;& P_{\mathrm{t}} + G_{\mathrm{t}} + G_{\mathrm{r}} + \sigma_{\mathrm{s}} + 20\log_{10}(\lambda)- 30\log_{10}(4\pi)\\
&- 40\log_{10}(R_{\mathrm{t\rightarrow s}})  + 20\log{10}(a).
\end{aligned}
\label{eq:bistatic_radar_a}
\end{equation}

We distinguish the following cases for the range ratio $a$:

\begin{equation}
a =
\begin{cases}
    > 1 & \text{$R_{\mathrm{t\rightarrow s}} > R_{\mathrm{s\rightarrow r}}$} \\
    = 1 & \text{$R_{\mathrm{t\rightarrow s}} = R_{\mathrm{s\rightarrow r}}$ (monostatic case)}. \\
    < 1 & \text{$R_{\mathrm{t\rightarrow s}} < R_{\mathrm{s\rightarrow r}}$}
\end{cases}
\end{equation}

For our deployment scenario in \cref{sec:deployment} we now compare a monostatic and a multistatic radar configuration side by side assuming the following basic system parameters:
\begin{table}[htbp]
    \centering
    \caption{System simulation parameters.}
    \label{tab:systemparameter}
    \renewcommand{\arraystretch}{1.3}
    \begin{tcolorbox}[
        enhanced,
        hbox,
        clip upper,                  
        boxrule=\fboxrule,           
        colframe=gray!50,            
        colback=gray!10,             
        arc=5pt,                     
        boxsep=0pt, left=0pt, right=0pt, top=0pt, bottom=0pt 
    ]
    \begin{tabular}{>{\columncolor{gray!15}}c r r}
        \rowcolor{gray!30}
        \textbf{Parameter} & \textbf{Monostatic} & \textbf{Multistatic} \\
        \hline
        $P_{\mathrm{t}}$         & \multicolumn{2}{c}{\SI{44}{\dB m}}\\
        $G_{\mathrm{t}}$         & \multicolumn{2}{c}{\SI{16}{\dB i}}\\
        $G_{\mathrm{r}}$         & \SI{16}{\dB i}                      & \SI{10}{\dB i} \\
        $\sigma_{\mathrm{s}}$    & \multicolumn{2}{c}{\SI{-10}{\dB sm}}\\
        $\lambda$                & \multicolumn{2}{c}{\SI{0.4283}{\metre} / @\SI{700}{\MHz}} \\
    \end{tabular}
    \end{tcolorbox}
    \vspace{-15pt}
\end{table}

The simulation results for both setups are illustrated in \cref{fig:recpower}. While the monostatic power follows \cref{eq:bistatic_radar}, the multistatic scenario evaluates three separate bistatic configurations, combining them into a joint dataset based on a worst-case selection of the weakest power level.

It can be seen clearly that the received power within the area to be protected is consistently higher for the multistatic configuration.
This effect is driven by the shorter receiver ranges \(R_{\mathrm{s\rightarrow r}}\) compared to the monostatic baseline.

\begin{figure}
    \centering
    \includegraphics[width=.9\linewidth,trim=0 0 2 2, clip,bgcolor=gray!10,rndcorners=5,rndframe={color=gray!50, width=\fboxrule, sep=\fboxsep}{5}]{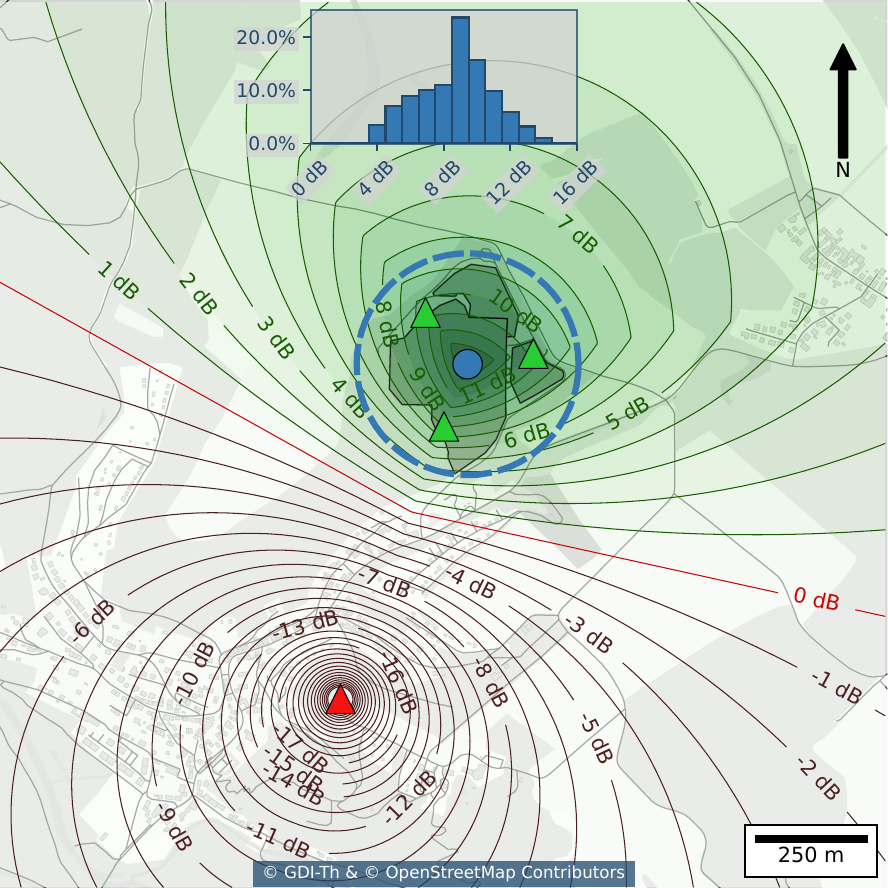}
    \caption{Spatial distribution of the range ratio \(a\) within the deployment scenario (represented by isolines), with the corresponding distribution across the \gls{ps} illustrated as an inset histogram.}
    \label{fig:multistaticgain}
    \vspace{-15pt}
\end{figure}

\begin{figure}[b]
    \centering
    \includegraphics[width=.9\linewidth,trim=-20 -1 -20 40, clip,bgcolor=gray!10,rndcorners=5,rndframe={color=gray!50, width=\fboxrule, sep=\fboxsep}{5}]{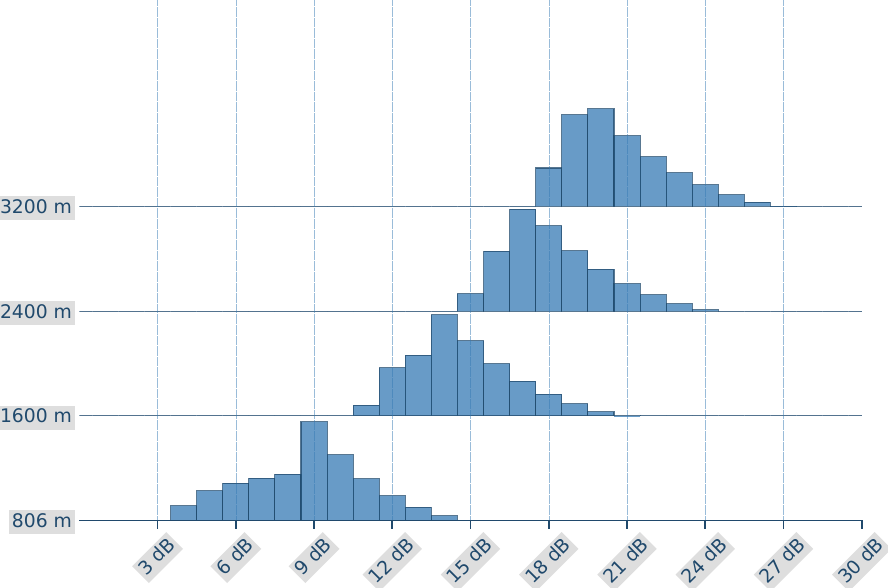}
    \caption{Range ratio distributions across the \gls{ps} illustrated as histograms depended on the distance of the \gls{bs} (\SI{806}{\metre} corresponds to \textit{real} distance in the deployment scenario, see \cref{fig:multistaticgain}).}
    \label{fig:gains}
    \vspace{-15pt}
\end{figure}

In terms of the previously introduced range ratio given in \cref{eq:rangeratio}, this geometric advantage corresponds to an increased value of \(a > 1\), which scale-up the received power via the \(+20\log_{10}(a)\) term in \cref{eq:bistatic_radar_a}.
\Cref{fig:multistaticgain} illustrates the spatial distribution of the range ratio \(a\).
Following the worst-case approach, the three separate bistatic configurations are combined into a joint dataset based on the minimum power level selection.
The distribution of the range ratio \(a\) illustrated in \cref{fig:multistaticgain} is a direct function of the distance between the illuminating \gls{bs} and the deployment's \gls{ps}.
A comparison with \cref{fig:gains} highlights this trend, showing that the bistatic gain intensifies as the hypothetical distances increase.

\begin{figure}
    \centering
    \includegraphics[width=.9\linewidth,trim=0 0 2 2, clip,bgcolor=gray!10,rndcorners=5,rndframe={color=gray!50, width=\fboxrule, sep=\fboxsep}{5}]{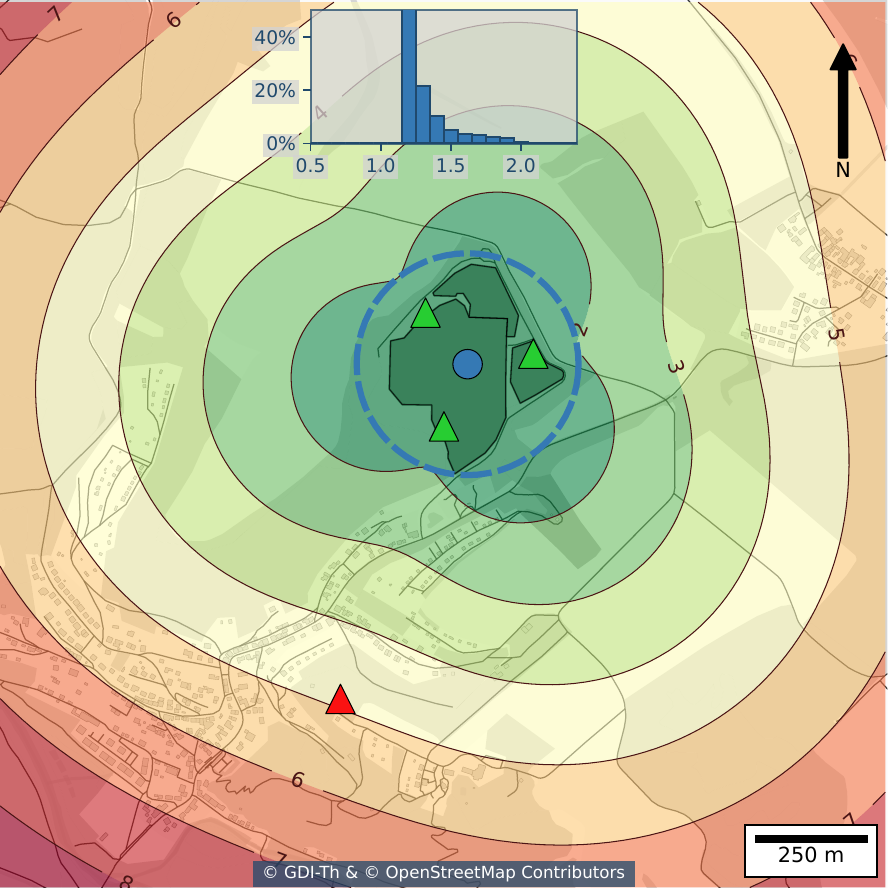}
    \caption{\Gls{hdop} within the deployment scenario (represented by isolines), with the corresponding distribution across the \gls{ps} illustrated as an inset histogram.}
    \label{fig:hdop}
    \vspace{-15pt}
\end{figure}

While \cref{eq:bistatic_radar_a} describes the received power resulting from propagation in free space, the radar performance further depends on the receiver's noise figure, the correlation matched filter correlation gain, and the coherent integration time.
The resulting \gls{snr} determines not only the target detection probability.
It also determines the variance of the bistatic range estimate \(\hat{\tau}\) according to the \gls{crlb} with \gls{ofdm} bandwidth \(B\)

\begin{equation}
    \operatorname{var}(\hat{\tau}) \geq \frac{3}{2\pi^2 \, B^2 \, \text{SNR}}
    \label{eq:CRLB}    
\end{equation}
From the measured bistatic \gls{etof} and \gls{edoppler} values the 3D dynamic state vector in Euclidean coordinates is calculated as described in \cite{Thoma2026MSISAC}.
The resulting variance of position and speed scales with a number that depends on the spatial distribution of the sniffers according to an error propagation mechanism known from satellite navigation as \gls{gdop} \cite{Langley1999DOP}.
\Cref{fig:hdop} shows the rms scaling reduced to the two horizontal dimensions $x,y$ as an example. The rms scaling factor is therefore called \gls{hdop} with position estimate $\hat{x}$, $\hat{y}$:

\begin{equation}
    \text{HDoP} = \sqrt{ var(\hat{x}) + var(\hat{y})}.
\end{equation}

\section{Results and Discussion}
\label{sec:results}
The distributed \gls{msue}-\gls{isac} approach which uses dedicated sniffers units deployed close to the target area promises a high SNR gain compared to the common (quasi) monostatic setup especially when the area to be protected is well defined, finite, and in some distance to the illuminating \gls{gnb}. The reason for this is that the bistatic radar equation then reduces to the case where only the \gls{tx} to target path dominates the end-to-end attenuation and the other path contributes less. Another advantage is that the direct feed-through from the transmitter to the receiver is drastically reduced. While strong \gls{tx} to \gls{rx} interference is often complained about in the quasi-monostatic \gls{ss}-\gls{isac} approach since it may saturate the receiver, the direct link in the distributed bistatic case is important, since it serves as a reference for differential \gls{tof} and Doppler measurement, thereby reducing synchronization issues a lot. Therefore, the sniffers should be placed to have LOS to the \gls{gnb}. The spatial distribution of the sensors (typically three or more) around the target area allows for the acquisition of a sufficient number of geometric degrees of freedom to enable the estimation of the complete 3D dynamic target state vector (position and speed) by multilateration.
Finally, we would like to point out that the \gls{snr} improvement we achieve by properly positioning the sniffer \glspl{ue} not only extends the detection range. Perhaps even more importantly, we can reduce the estimation variance within a given target distance according to the \gls{crlb}, \cref{eq:CRLB}, which improves localization accuracy.   
The \gls{snr} influence to \gls{crlb} also explains that lower operating frequencies (e.g. \gls{fr1} low- and midband) with their lower bandwidths are not as disadvantageous as often presumed since that is partly compensated by higher \gls{snr} (for a given distance). In addition, the lower antenna directivity comes as an advantage because target beam search is relaxed. 

As for the limited target resolution relative to the direct \gls{los} link between the \gls{tx} and \gls{rx}, as well as with respect to the resolution of multiple targets, we suggest that resolution is achieved not only in the distance dimension but also in the Doppler dimension. And since the \gls{los} response is well known (actually we have estimated the correlation reference), we can remove this contribution even if it overlaps to the target response by some kind of interference cancellation procedure as long as the receiver is in its linear operation mode. See \cite{Thoma2026MSISAC} for some deeper discussion of model based high resolution in joint range and Doppler dimension.   

Finally, we must point out that the performance evaluation of our proposal for a distributed \gls{msue}-\gls{isac} presented in this article is conducted solely on a relative basis, which consists in comparison to the (quasi) monostatic case. If we assume that the same \gls{gnb} parameters (frequency, bandwidth, \gls{eirp}) are applied in both cases and if the receiver achieves the same matched filter processing gain and the same coherent radar integration time, the \gls{snr} gain indicated by the range ratio $a$ in \cref{eq:bistatic_radar_a} and \cref{fig:multistaticgain,fig:gains} results only from the choice of the positions of the sniffer \glspl{ue}.
Some difference still may remain due to the size of the antenna arrays on the sniffers in both cases.
The antenna array of a sniffer located near the \gls{gnb} may have higher gain than that of the sniffer located near the target, since the latter requires full \SI{360}{\degree} coverage.
On the other hand, since we have three or more sniffers around, we can achieve another \gls{lms} estimation gain by fusing the measurement from those sniffers.

\section{Outlook and Conclusion}
\label{sec:conclusion}
We have proposed a distributed \gls{ms}-\gls{isac} architecture consisting of multiple dedicated sniffer units connected to the base station via the standard multiuser \glspl{dl} / \glspl{ul}.
The advantages include extended coverage and low \gls{gdop} value within the give service area.
Another asset is that the \gls{bs}’s \gls{ran} interface would not require any significant changes to the hardware or software, which could lower the barriers to acceptance for mobile network operators.
The sniffer works similarly to the well-known passive radar.
But according to \gls{cpcl}, it can interact with the base station at the application software level. 

We have shown that, provided a suitable geometric arrangement of the sniffers is possible, the \gls{msue}-\gls{isac} approach outperforms the conventional (quasi-)monostatic approach in several respects.
However, one might argue that the coverage area is possibly too limited and that the concept of using dedicated sniffers is not a suitable approach for achieving comprehensive wide area coverage.
As a solution to this dilemma, we propose rethinking the well-known concept of ''umbrella cells” and adapting it to \gls{isac}. If a macro cell - especially in the lower \gls{fr1} band - covers a large area, there will usually be several small-cell base stations from the same operator within reach.
In that case, the respective mobile network operator can reuse its own sites to host dedicated sniffer units, which requires installing an additional \gls{fr1} receiving antenna and the corresponding sniffer receiver.
This configuration works the same way as the \gls{msue}-\gls{isac} sniffer,  except that it does not need a full \gls{ul} / \gls{dl} for data reporting since we have the network backbone at hand which would be enough for compressed data reporting. Wide area coverage is supported by the wider service range at lower \gls{fr1} frequencies while higher resolution may be achieved at local sensing hot spots with the \gls{msue}-\gls{isac} approach at \gls{fr1} midband or prospective \gls{fr3} frequencies. This creates a comprehensive, hierarchically structured \gls{ue} monitoring network that can be gradually put into operation and expanded.  

An even more general and powerful architecture for infrastructure-oriented \gls{ms}-\glspl{isac} can be achieved when a \gls{comp} RAN network is available to serve as a MS ISAC cluster, for example, within a campus network that covers a large industrial area requiring protection. In this case, we have several \glspl{ru} connected to a \gls{du} and the collocated \gls{bbu} pool by a fiber-optic fronthaul link while the functional split follows the \gls{openran} convention. We refer to this approach as \gls{msru}-\gls{isac}. This architecture approach allows fast resource allocation and \glspl{ru} access coordination. It enables a full meshed \gls{ms}-\gls{mimo} architecture as each RU can act as an illuminator while all others serve as sensors \cite{Thoma2026MSISAC}.

%
%
%

%
\subsection*{Acknowledgments}
This work is partially supported by the the Federal Ministry of Research, Technology and Space of Germany (BMFTR) in the projects SENSATION (grant number: 16KIS2531) and Open6GHub+ (grant number: 16KIS2406) as well as by the Deutsche Forschungsgemeinschaft (DFG) under the project JCRS CoMP (grant number: TH 494/35-1).
%
%
%
%

\printbibliography

@ARTICLE{Keskin2025FundTradoffMonoStaISAC,
  author={Keskin, Musa Furkan and Mojahedian, Mohammad Mahdi and Lacruz, Jesus O. and Marcus, Carina and Eriksson, Olof and Giorgetti, Andrea and Widmer, Joerg and Wymeersch, Henk},
  journal={IEEE Transactions on Wireless Communications}, 
  title={Fundamental Trade-Offs in Monostatic ISAC: A Holistic Investigation Toward 6G}, 
  year={2025},
  volume={24},
  number={9},
  pages={7856-7873},
  doi={10.1109/TWC.2025.3563197}
}

@misc{Blandino2026MonoStaSenUAV,
      title={Evaluation of gNB Monostatic Sensing for UAV Use Case}, 
      author={Steve Blandino and Neeraj Varshney and Jian Wang and Jack Chuang and Camillo Gentile and Nada Golmie},
      year={2026},
      eprint={2604.02205},
      archivePrefix={arXiv},
      primaryClass={eess.SP},
      url={https://arxiv.org/abs/2604.02205}, 
}

@article{Langley1999DOP,
  author  = {Richard B. Langley},
  title   = {Dilution of Precision},
  journal = {GPS World},
  volume  = {10},
  number  = {5},
  pages   = {52--59},
  year    = {1999},
  month   = may
}

@article{Shatov2024JRC,
  author  = {Victor Shatov and Benjamin Nuss and Steffen Schieler and
             Pradyumna Kumar Bishoyi and Lara Wimmer and Maximilian L\"ubke and
             Navid Keshtiarast and Christoph Fischer and Daniel Lindenschmitt and
             Benedikt Geiger and Reiner S. Thom\"a and Amina Fellan and
             Laurent Schmalen and Marina Petrova and Hans D. Schotten and
             Norman Franchi},
  title   = {Joint Radar and Communications: Architectures, Use Cases,
             Aspects of Radio Access, Signal Processing, and Hardware},
  journal = {IEEE Access},
  volume  = {12},
  pages   = {47888--47914},
  year    = {2024},
  doi     = {10.1109/ACCESS.2024.3383771}
}

@article{Thoma2019CPCL,
  author  = {Reiner S. Thom\"a and Carsten Andrich and Giovanni Del Galdo and
             Michael D\"obereiner and Matthias A. Hein and Martin K\"aske and
             G\"unter Sch\"afer and Steffen Schieler and Christian Schneider and
             Andreas Schwind and Philip Wendland},
  title   = {Cooperative Passive Coherent Location: A Promising 5G Service to Support Road Safety},
  journal = {IEEE Communications Magazine},
  volume  = {57},
  number  = {9},
  pages   = {86--92},
  month   = sep,
  year    = {2019},
  doi     = {10.1109/MCOM.001.1800242}
}

@article{Thoma2026MSISAC,
  author  = {Reiner S. Thom\"a and Carsten Andrich and Michael D\"obereiner and
             Reza Faramarzahangari and Jonas Gedschold and
             Marc Francisco Colaco Miranda and Saw James Myint and
             Steffen Schieler and Christian Schneider and Sebastian Semper and
             Carsten Smeenk and Gerd Sommerkorn and Zhixiang Zhao},
  title   = {Distributed Multisensor ISAC},
  journal = {npj Wireless Technology},
  volume  = {2},
  number  = {1},
  year    = {2026},
  doi     = {10.1038/s44459-026-00041-2}
}

@article{Saur2026UAVDetection,
  author  = {Stefan Saur and Michael Doll and Alexander Grudnitsky and
             Stefano Mandelli and Luca Giroto and Markus Henninger and
             Thomas Wild},
  title   = {Reliable UAV Detection with ISAC},
  journal = {arXiv preprint arXiv:2605.23561},
  year    = {2026},
  doi     = {10.48550/arXiv.2605.23561}
}
\end{document}